# Usage History of Scientific Literature: Nature Metrics and Metrics of *Nature* Publications

Xianwen Wang[1,2]*, Wenli Mao[1†], Shenmeng Xu[3†] & Chunbo Zhang[1]

[1]WISE Lab, School of Public Administration and Law, Dalian University of Technology, Dalian 116085, China.
[2]DUT- Drexel Joint Institute for the Study of Knowledge Visualization and Scientific Discovery, Dalian University of Technology, Dalian 116085, China.
[3]School of Information and Library Science, University of North Carolina at Chapel Hill, Chapel Hill, NC 27599-3360.

† These authors contributed equally as second authors.
* Corresponding author. Tel.: +86 411 847 072 92-23; fax: +86 411 847 060 82.
Email address: xianwenwang@dlut.edu.cn; xwang.dlut@gmail.com

**Abstract:** In this study, we analyze the dynamic usage history of Nature publications over time using *Nature* metrics data. We conduct analysis from two perspectives. On the one hand, we examine how long it takes before the articles' downloads reach 50%/80% of the total; on the other hand, we compare the percentage of total downloads in 7 days, 30 days, and 100 days after publication. In general, papers are downloaded most frequently within a short time period right after their publication. And we find that compared with Non-Open Access papers, readers' attention on Open Access publications are more enduring. Based on the usage data of a newly published paper, regression analysis could predict the future expected total usage counts.

***Keywords***: altmetrics; article-level metrics; download; Nature metrics; page view; usage data

## 1. Introduction

Traditional metrics of scientific articles were mostly based on publication data. Nevertheless, metrics based on usage data are increasingly being used in recent years. A variety of usage metrics are applied in scientometrics studies, for instance, research evaluation (Davis et al. 2008; Davis and Solla 2003), impact assessment (Brody et al. 2006; Davis et al. 2008; Shuai et al. 2012), and user behavior study (Davis and Price 2006; Davis and Solla 2003).

Scientific publishers record and store usage information of each article, and sometimes they report this information to editors or editorial board (Thelwall 2012). However, this kind of usage data is rarely made public. On most of the mainstream publishing platforms, it's very difficult for people to know how many times one paper has been downloaded. However, in recent years, usage data for readers gradually drew attention from publishers. Here are some of the few examples, as Table 1 shows.

Table 1 List of some publishers' usage statistic tool

| Publisher/Digital library | Usage statistic tool | Summary |
|---|---|---|
| ACM DL | Bibliometrics | Downloads (6 Weeks/12 Months), cumulative downloads for each article and journal |
| ADS Abstract Service | Reads history | For each paper, a "read" is counted if an ADS user runs a search in our system and then requests to either view the paper's full bibliographic record or download the full-text. |
| Elsevier | Top 25 Hottest | 25 most-read articles during the prior three |

|  | Articles | months |
|---|---|---|
| *Wiley* | Most Accessed | 10 most-accessed articles in the prior month |
| *Nature* | Top content<br>Most emailed articles<br>Most read articles | This list of most-read articles is created by calculating article views for the previous four weeks (28 days). It is refreshed daily. |
| *Nature* | Nature metrics | Daily page views counts for each research paper |
| *PLOS* | Metrics | Article views for each paper in each month |
| *Sage* | Most Read | 50 most-read articles, update monthly |
| *Springer* | Most downloaded articles | 5 most-read articles during the prior 7/30/90 days |
| *Springer* | Realtime.springer.com | The *Feed* tool shows which papers are being downloaded. |
| *Taylor & Francis* | Article views | Article usage statistics combine cumulative total PDF downloads and full-text HTML views from publication date |
| *Taylor & Francis* | Most read articles | 20 most read articles, updated every 24 hours based on user behavior. |

Among these usage statistic tools, most of them (including Elsevier, Springer, Wiley, etc.) only report the most downloaded articles, but the usage information for each article is not available. Some publishers and digital libraries provide article-level usage data, however, they are updated slowly. For example, the download counts displayed in *ACM DL* are usually 1-2 weeks behind the current date. Nevertheless, *Nature*, *Taylor & Francis*, and *PLOS* update their article usage statistics more timely, which are on a daily basis. Besides the total article views, PLOS also provides data month-by-month. And *Nature metrics* reports detailed cumulative page views every day after the publication of each paper. Another example is the *Realtime* platform of Springer, on which the *Feed* tool shows which papers are being downloaded right now.

## 2. Related studies

*Usage metrics*
Digital libraries have massive server logs of user's retrieval requests, which made it possible to conduct "retrieval analysis" or "download analysis" to study the retrieval habits of users, and to assess the impact of scientific work based on the downloads (Bollen and Luce 2002; Kaplan and Nelson 2000; Marek and Valauskas 2002).

Taking the NASA Astrophysics Data System (ADS) Abstract Service as their research object, Kurtz et al. did a series of studies about the readership logs (Kurtz et al. 2005a), on readership and citation (Kurtz et al. 2005b; Henneken et al. 2010), usage patterns (Henneken et al. 2009), etc. They conclud that "We now know how many times an article is read, where the reader is from, and "who" (as a unique cookie identifier, not as a name, which remains anonymous) the reader is. The existence of this information has great implications for the future of information retrieval and bibliometrics." (Kurtz et al. 2005b).
Moreover, some previous studies show significant correlation between the early usage statistic and later citation impact (Brody et al. 2006; Shuai et al. 2012).
In these studies, static usage data like the cumulative downloads for an article are collected. Unlike the static usage data used in previous studies, dynamic real-time usage data collected from *realtime.springer.com* can be used to make more detailed analysis on how a scientific paper is being used after publication. In one of the studies that we conducted, we examined at what time people download paper from Springer. Converting the time data according to the time zones where the request originated, we were able to see how hard scientists work overall (Wang et al. 2013; Wang et al. 2012b). In another study, we recorded and analyzed the papers being downloaded to estimate what kind of research scientists are doing (Wang et al. 2012a).

*Altmetrics*
As the development of social network, scientific papers are producing increasing impact on web environment. "To develop alternative methods for scholars or research institutions, authors, journal editors, and academic publishers to use Web sources for additional citations to their work", a new combined Integrated Online Impact (IOI) indicator is introduced (Kousha et al. 2010).
Altmetrics is new metrics based on the social web, which aims to make a real-time analysis of the scholarly impact of articles (J. Priem et al. 2010; Jason Priem and Hemminger 2010). Unlike traditional and classic scientometrics indicators of impact assessments, which only focus on citation counts, altmetrics captures various aspects of the impact of a paper, including article views/downloads, citations, mentions in social/blog/news media, and other tag data in academic social bookmarks such as Mendely, CiteUlike, F1000Prime, etc (Galligan and Dyas-Correia 2013; Lin and Fenner 2013; ImpactStory 2012).

## 3. Data and Methods

As of October, 2012, *Nature* began to launch a real-time online count of article-level metrics for its published research papers published on or after 1 January 2012 (Nature, 2012). *Nature Metrics* provides citation data (WOS, CrossRef and Scopus), online attention data (Altmetric score) and usage statistics (page views) for every research article of *Nature*, as Figure 1 shows. 20 NPG (Nature Publishing Group) journals published on *nature.com* are included. This count provides an alternative measure to track research impact and evaluate scientific output.

Unlike merely gross usage statistics provided by other publishers, the "page views" not only covers the cumulative count of full-text article views that includes HTML views and PDF downloads, but also gives daily counts since the publication date. According to the official statement of *Nature*, "the page views data is available 48 hours after online publication and is updated daily."

For the page views, HTML views and PDF downloads are treated as the same. However, these two counts could be different. For example, PDF tends to be the preferred format if researchers want to print the article or just save in hard discs for later study. If one paper was downloaded as PDF, it tends to be seen as more valuable than another paper which was only viewed in the browser. It's worth mentioning here that PLOS reports its usage data in 3 different formats (HTML Page views, PDF Downloads and XML Downloads).

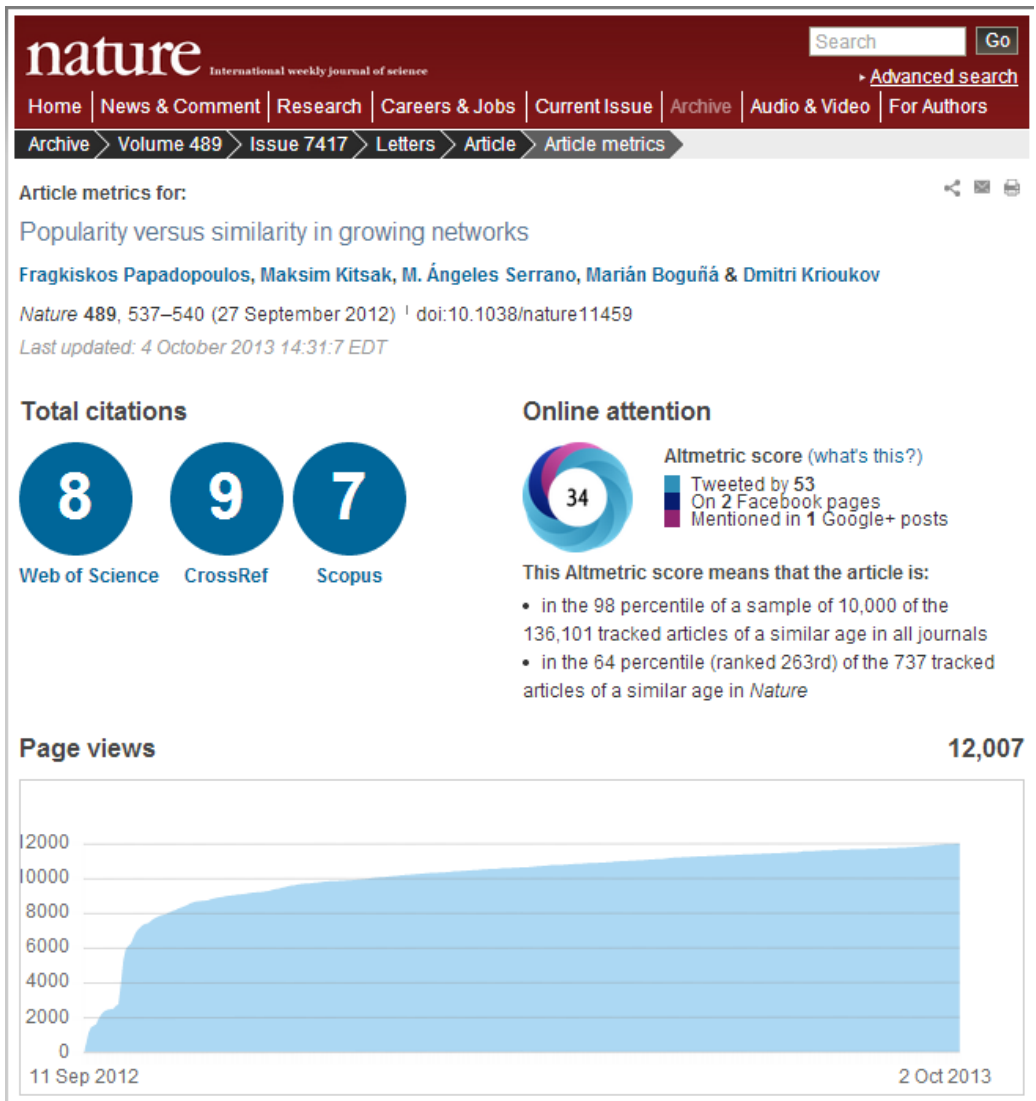

Figure 1 Article metrics for an example of *Nature* publication

*Nature* published 51 issues in 2012 (from volume 7379 to volume 7429). Among all the 1124 research publications available for *Nature metrics*, there are 159 Articles, 665 Letters, 11 Reviews, 252 Correspondences, 24 Brief Communication Arisings, 7 Perspectives, and 6 Insights.

In order to guarantee a time span long enough for each sample, here we study all the articles/letters published before September 1st, 2012. Moreover, considering the significantly distinct downloading patterns of pre-dated publications and instant publications, we exclude the items of which the online publication date is relatively long (2 days or longer) before the issue date. In other words, in this research, the online publication date of all the samples is in accordance with the issue date. Finally, 185 samples, involving 35 articles and 150 letters, are selected as our research objects.

In this study, only the indicator of "page views" is used. We trace and record the everyday "page views" data of our samples. For instance, the paper of 10.1038/nature10666 was published on January 4th, 2012, so we set the day as Day 0. Accordingly, January 5th is Day 1, and January 6th is Day 2, and so on.

# 4. Results

**3.1 Cumulative counts of page views of nature articles**

Figure 2-a and Figure 2-b illustrate the cumulative counts of page reviews of *Nature* articles. X-coordinate indicates number of days after the publication date, while y-coordinate indicates cumulative page views counts. Articles and letters are displayed in different colors. For Open Access (OA) articles/letters, the curves are bold and darker. The smaller range of y-coordinate in Figure 2-b magnifies the curves of non-OA articles. The color schemes are the same. Figure 2-c shows the comparison of page views of OA articles/letters and the mean and median value of all papers.

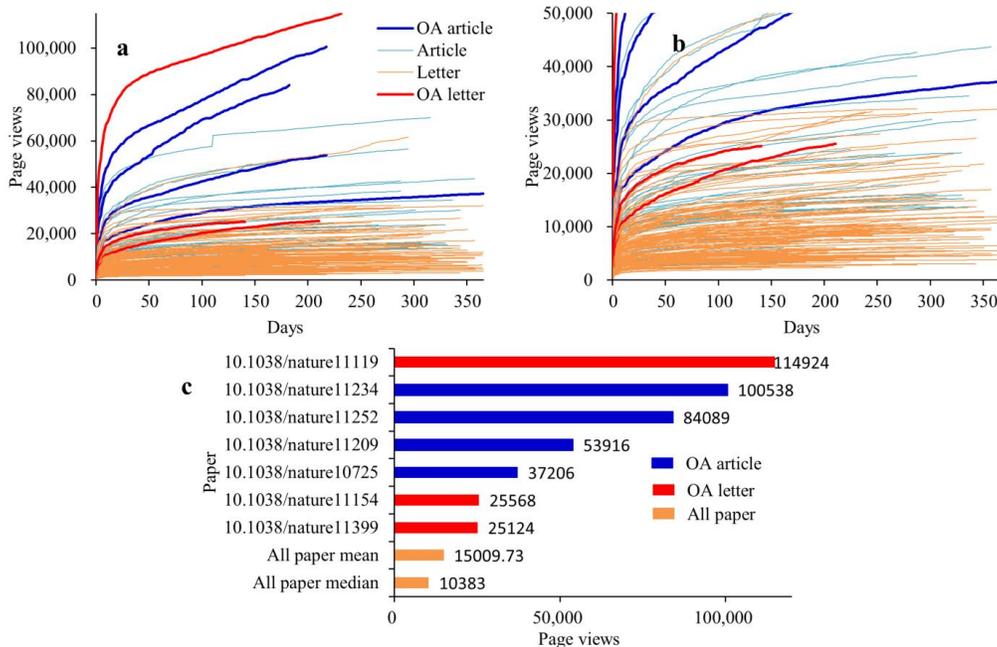

Figure 2 Page views of Nature papers published before 2012-09-01.

As Figure 2 displays, among the 4 Open Access (OA) articles, 2 of them (10.1038/nature11234 and 10.1038/nature11252) have relatively high page views. The other 2 OA articles are also reviewed more often than ordinary articles. Meanwhile, the page views of one Open Access letter (10.1038/nature11119) is extraordinarily high, dwarfing the other 2. For all the 185 articles/letters, the number of average page views is 15009.73, and the median value is 10383. Notably, for the 7 Open Access articles/letters, the maximum value is 114924. OA papers have a significantly higher value of page views than those not open.

**3.2 Time before page views reach 50% / 80% of total**

We continue to analyze the trends of the page views over time. As we calculated, for these 185 articles and letters, it takes averagely 7.92 days to reach 50% of the total page views. The median of our samples is 7 days, which is quite coincident with the weekly publishing periodicity of *Nature*. The papers with the fastest page views growth rate (10.1038/nature10906, 10.1038/nature11084, and 10.1038/nature11281) were viewed half of the total times only within 2 days, while the value for the slowest paper (10.1038/nature10932) is 27 days.

The growth of page views is tend to be affected by information news worthiness and competition of new information, which will restrict and reduce the growth rate (Wei, Bu, & Liang, 2012). Calculating the value for 80% of the total page views, we find that it takes much longer

(63.14 days) than to reach 50% of the total views. And as shown in Table 2, the median is 59 days. In addition, the paper of 10.1038/nature10906 got 80% of its page reviews within 9 days after publication, while it took 168 days for 10.1038/nature10927.

Table 2 Statistics of days before page views reach 50%/80% of the total

|  | 50% of total page views | 80% of total page views |
| --- | --- | --- |
| Mean days | 7.92 | 63.14 |
| Median days | 7 | 59 |
| Minimum days | 2 | 9 |
| Maximum days | 27 | 168 |
| Samples | 185 | 185 |

Figure 3-a and 3-b show the number and percentage of articles/letters which gain 50% (a) and 80% (b) of its total page views in certain time periods. X-coordinate denotes time after the publication date. Blue bars show number of papers which attain 50%/80% page views in different time periods. Dotted orange line shows the cumulative percentage. Figure 3-c compares the time before 50%/80% views of the 7 Open Access (OA) articles/letters and the mean and median time of all 185 articles/letters.

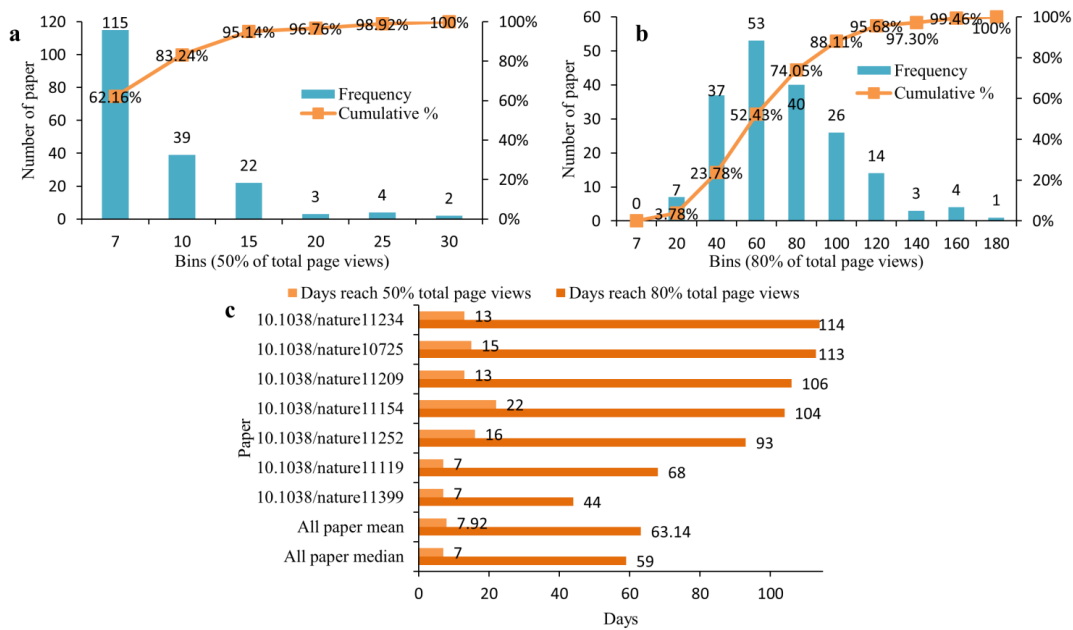

Figure 3 Statistics of time before their page views reach 50%/80% of the total

As is displayed in Figure 3, there are 115 articles/letters reaching 50% of the total page views within 7 days. That accounts for 62.16% of the 185 papers. Cumulatively, 83.24% of all the papers gain 50% of their page views within 10 days, and 95.14% of them gain half of the reviews within 15 days. Meanwhile, only 7 papers, accounting for 3.78% of the total papers, can reach that within 20 days. It takes 23.78% of all the papers 40 days, 52.34% of them 60 days, and 88.11% 100 days.

Notably, most OA papers need a longer time window to get 50%/80% of all their page views than the mean and median value. For instance, it takes 22 days for 10.1038/nature11154 and 16 days for 10.1038/nature11252 to reach 50% of total page views. And 4 of the 7 OA papers reach 80% of the total page views after more than 100 days.

### 3.3 Page views in certain periods of time after publication

Furthermore, we calculate the page views in certain periods of time after publication. Here we set the time nodes as 7 days, 30 days, and 100 days. After 7 days, the paper with the highest page view percentage (10.1038/nature10906) gained 77.23% of its total counts, while the "slowest" paper (10.1038/nature10932) only gained 33.06% of its views. The median value is 52.77%. After 30 days, the paper of 10.1038/nature10906 reaches as high as 91.73% of its total page views, when 10.1038/nature10932keeps the lowest percentage (52.64%). And the median value is 72.36%. 100 days after publication, the paper of 10.1038/nature11340 took the place of 10.1038/nature10906, with a percentage of 96.61% of its total page views, while the paper of 10.1038/nature10932 still keeps the lowest percentage with a percentage of 71.44%. The median value here is 86.89%.

We see from the detailed statistics in Table 4 that generally, papers gain above 52% of their total page views within 7 days after publication. After about one month, they gain above 72% of the total counts. And the number would excess 86% within 100 days.

Table 3 Percentage of total page views in certain periods of time after publication

|        | 7 days  | 30 days | 100 days |
|--------|---------|---------|----------|
| Max    | 77.23%  | 91.73%  | 96.61%   |
| Median | 52.77%  | 72.36%  | 86.89%   |
| Min    | 33.06%  | 52.63%  | 71.44%   |

Figure 4a-c illustrate the number and percentage of articles/letters which attain certain percentage of the total page views in 7days(a), 30 days(b), and 100 days(c) respectively after publication. X-coordinate denotes the percentage of total page views papers attain. Blue bars show number of papers which gain the corresponding percentage of total views. Dotted orange lines show the cumulative percentage. Figure 4d-f show the comparisons of the percentages of the 7 Open Access articles/letters' page views and the mean and median value of all papers in these 3 time periods.

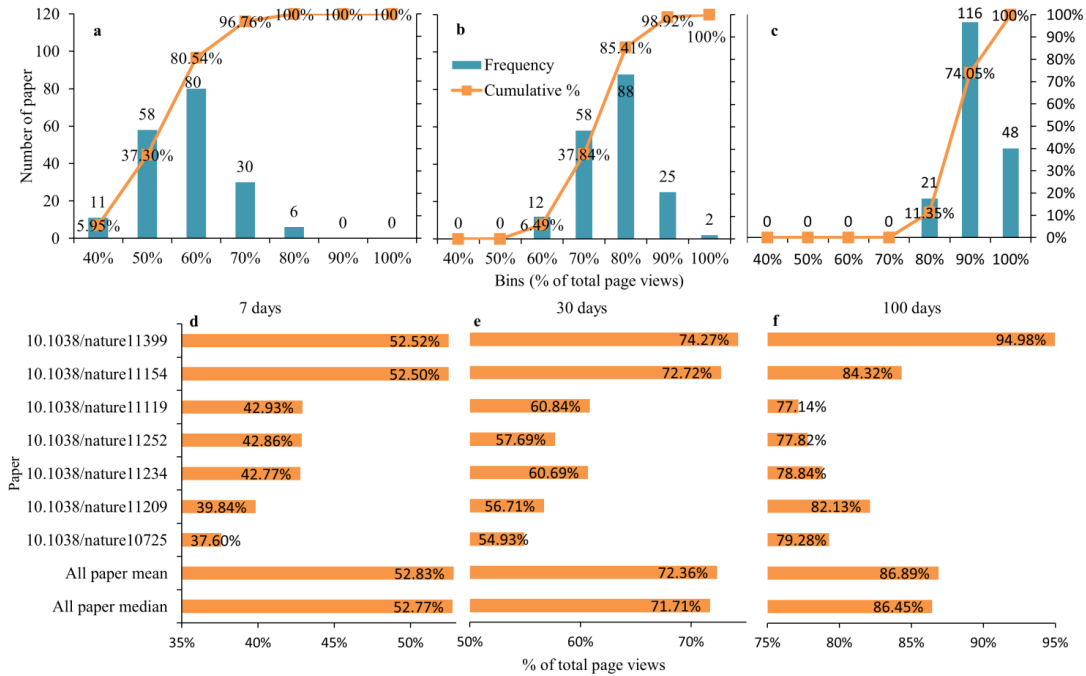

Figure 4 Statistics of the percentage of total page views in 7/30/100 days after publication

## 3.4 Regression analysis

As Figure 5a, b show, in the initial stage after publication, the line of page views follows logarithmic distribution. Nevertheless, in the later stage (as is indicated by the right-most part of the curve), the value of page views fit into liner distribution.

In Figure 5, Y-coordinate denotes the articles' total page views in 100 days after publication, and the X-coordinate in Figure 5a, b denotes the corresponding total page views in 7 and 15 days, respectively. The data in both panels of Figure 5 fit well into liner distribution.

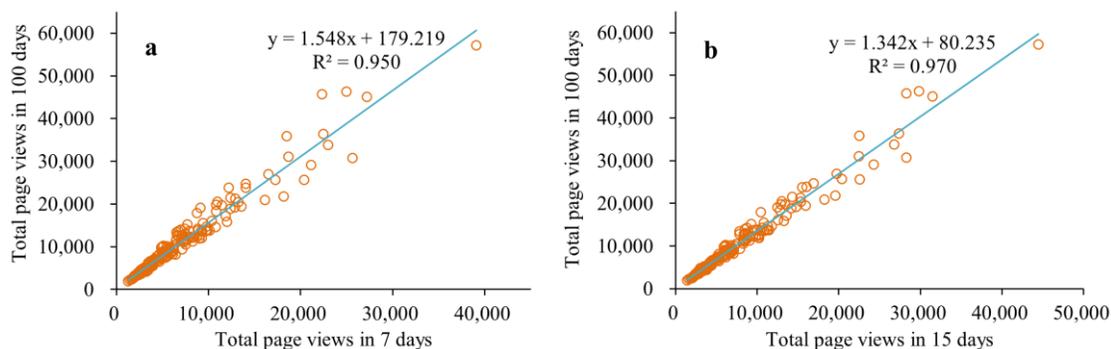

Figure 5 Scatter plot of total page views in 100 days and 7/15 days

As a result, given the value of the starting point of the liner distribution, the value of other right points could be estimated using a unary linear regression model.

$$y=a+b(x) \quad (1)$$

Where $y$ is the estimated accumulated page views in a long time period, e.g. 100 days, and $x$ represents the value of the starting point of the liner distribution, i.e., the page views in a short time here.

Two possible starting points could be considered for the regression analysis, which are day 7 and day 15. What we want to do is to draw the prediction curve, so that given the actual page views of a paper in 7/15 days, we can estimate the accumulated page views in 100 days. Here we exclude the 7 open access articles/letters, so 178 papers are selected as research samples.

Table 4 reports the regression results. For model 1, R-squared is 0.950, and for model 2, R-squared is 0.970. Accordingly, using the accumulated page views of day 15 as the starting point to estimate the future expected value is a better choice.

Table 4 Regression Results for accumulated page views in 100 days

|  | Model 1 | Model 2 |
|---|---|---|
| Constant | 179.219 (244.541) | 80.235 (189.678) |
| Day7 | 1.548*** (0.027) |  |
| Day15 |  | 1.342*** (0.018) |
| R-squared | 0.950 | 0.970 |
| Adjusted R-squared | 0.950 | 0.970 |
| No. observations | 178 | 178 |

Standard errors are reported in parentheses.

\* Significant at the 10% level.
\*\* Significant at the 5% level.
\*\*\* Significant at the 1% level.

## 5. Conclusion and discussion

Our study finds regular patterns from the page views data of *Nature metrics* over time. Papers tend to be viewed most frequently within a short time period after publication. Specifically for the articles/letters published on *Nature*, a majority of them, 62.16% approximately, are viewed more than half of their total times in the first week. Within the first month, all of the papers attain more than 50% of their page views, and in the first 2 months, 52.48% of the papers gain more than 80% of their total views. From another perspective, the page views number reaches more than 52% of the total in the first week and more than 72% in the first month, and then gradually grows to about 87% in 100 days. After one month, the growth rates sharply decline.

The attention history for Open Access articles is different from Non-Open Access ones. Compared with Non-OA paper, OA paper is more likely to obtain more page views. However, we find another interesting phenomenan. Compared with Non-OA paper, readers' attention on Open Access publications is more enduring. Even after a relatively long time of its publication, the OA papers still have a large number of downloads, but the downloads of Non-OA papers decrease much faster and more dramaticly.

Given the usage data of a newly published paper in a short time, e.g., 7 days/15 days for *Nature* papers, it is possible to predict future expected total usage counts.

Publication data and citation data have been dominating bibliometrics studies for a long time. As an emerging kind of data, usage data of electronic papers have great value and implications for the future of information retrieval and bibliometrics studies (Kurtz et al. 2005b). We are happy to see that more and more publishers and digital libraries are starting to report the usage data to public, among which *Nature* Metrics and Springer have become good examples in providing detailed usage data.

However, the format of usage data from different publishers are very different and hard to integrate for researchers. Another problem is that the dissimilarity of usage data types makes it impossible to comparatively study articles collected from different publishers. Accordingly, an industy standard should be made (Thelwall 2012).

There are limitations of our study. Firstly, using article usage data in scientometrics research needs to be scrutinized. For example, downloads may not have equal value, and papers may be downloaded but never read (Thelwall 2012). Also, sometimes, the download of an article may be intended for teaching purpose, rather than research purpose (Thelwall 2008). In addition, comparing to citation data or even online attention data from social media, usage data could be manipulated more easily, in direct or indirect ways. So, usage data should be interpreted with caution.

Secondly, in this paper, we only focus on the *Nature* publications. For other NPG journals published on nature.com, such as *Nature Chemistry*, *Nature Physics*, etc., the article metrics data are also available. Are the usage patterns similar? This is one of the questions we want to answer in the future.

Thirdly, the "page views" is the only indicator of *Nature* metrics in our study. In the future, we may include other indicators such as citation data and altmetric scores.

## Note

Shenmeng Xu was a master student in WISE Lab, Dalian University of Technology when the manuscript submitted to *Scientometrics*.

# Acknowledgements

The work was supported by the project of "National Natural Science Foundation of China" (61301227), "Social Science Foundation of China" (10CZX011) and the project of "Fundamental Research Funds for the Central Universities" (DUT12RW309).